\documentclass[nopubldata]{lpor2014}
\graphicspath{{figs/}}
\begin{document}
%
 \titlefigure{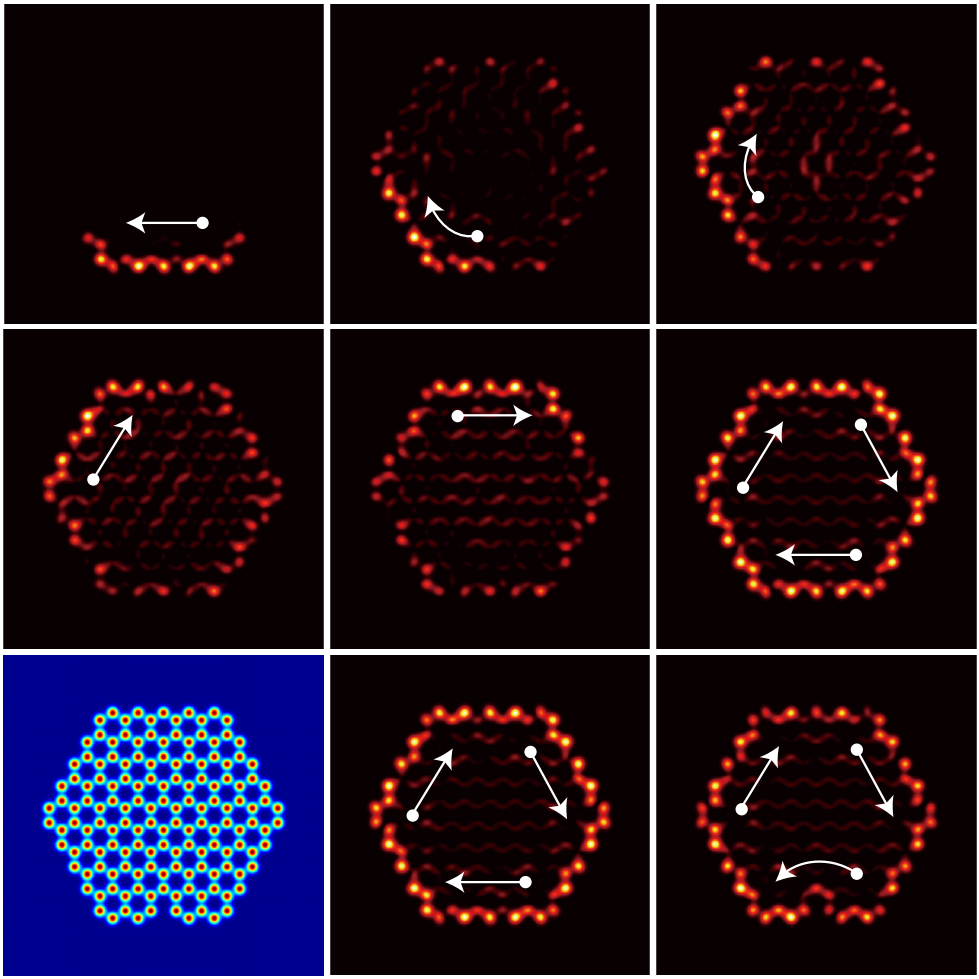}

\abstract{Photonic topological insulators supporting unidirectional topologically protected edge states represent attractive platform for realization of disorder- and backscattering-immune transport of edge excitations in both linear and nonlinear regimes. In many realizations of topological insulators structured periodic materials are used, since they may admit specific Dirac degeneracy in the spectrum, around which unidirectional edge states appear under the action of physical effects breaking time-reversal symmetry. While properties of the edge states at unclosed interfaces of two bulk media with different topology are known, the existence of the edge states in practical finite-dimensional topological insulators fully immersed in nontopological environment remains largely unexplored. In this work using as an example realistic polariton topological insulators built from small-size honeycomb arrays of microcavity pillars, we illustrate how topological properties of the system build up upon gradual increase of its dimensionality. To account for dissipative nature of polariton condensate forming in the array of microcavity pillars, we consider the impact of losses and resonant pump leading to rich bistability effects in this system. We describe the mechanism in accordance with which trivial-phase pump "selects" and excites specific nonlinear topological edge states circulating along the periphery of the structure in the azimuthal direction dictated by the direction of the external applied magnetic field. We also show the possibility of utilization of vortex pump with different topological charges for selective excitation of different edge currents.}

\title{Finite-dimensional bistable topological insulators: From small to large}
%
\titlerunning{}
\author{Weifeng Zhang\inst{1}, Xianfeng Chen\inst{1}, Yaroslav V. Kartashov\inst{2}, Dmitry V. Skryabin\inst{3}, and Fangwei Ye\inst{1,*}}
%
\authorrunning{}
%
\institute{%
School of Physics and Astronomy, Shanghai Jiao Tong University, Shanghai 200240, China
\and
Institute of Spectroscopy, Russian Academy of Sciences, Troitsk, Moscow Region, 108840, Russia
\and
Department of Physics, University of Bath, BA2 7AY, Bath, United Kingdom
}
%
\mail{\email{fangweiye@sjtu.edu.cn}}
%
%
\maketitle

\section{Introduction}
\label{sec:intro}
The phenomenology of topological insulators - materials with unusual physical properties, behaving like usual insulators in the bulk, but admitting conductance through topologically protected states appearing at their edges - have penetrated in diverse areas of science, ranging from solid-state physics, mechanics, physics of matter waves, to acoustics and optics. Considerable progress was made in the experimental realization of such states in all above mentioned areas, in particular in electronic systems (see reviews \cite{elect1,elect2} and references therein), where such states were introduced initially, {\color{red}mechanical systems, where tunable phononic topological insulators have been designed using geometrically induced effects of spin-orbit coupling \cite{mech1,mech2,mech3}}, acoustics \cite{acou1,acou2}, cold atoms in optical lattices \cite{cold1}, atomic Bose-Einstein condensates \cite{cond1}, and most notably in optical systems (see review \cite{topphot}), including gyromagnetic photonic crystals \cite{giro1,giro2}, semiconductor quantum wells \cite{well1}, arrays of coupled resonators \cite{res1,res2}, metamaterial superlattices \cite{meta1}, and helical waveguide arrays \cite{helix1,helix2,helix3,helix4,helix5}. Very recently, topological insulators were also predicted theoretically \cite{polar1,polar2,polar3} and realized experimentally \cite{polar4} in polaritonic systems. The representative advantage of optical and polaritonic systems is that they enable observation of rich interplay of topological and nonlinear effects. Moreover, in these systems diverse periodic structures can be utilized for construction of insulators with various geometries. Thus, modern technologies of polariton microcavity structuring \cite{polar5,polar6} allow creation of various periodic potential landscapes (see for example works \cite{polar7,polar8,polar9,polar10,polar11}) possessing characteristic Dirac degeneracies in the spectrum that are frequently required for formation of topological phases in such systems. Very high nonlinearities in polariton condensates allow observation of non-topological \cite{sol1,sol2,sol3} and potentially of topological \cite{sol4,sol5,sol6} solitons in such periodic potential landscapes. Nonlinear response may change circulation direction for topological currents \cite{sol7,sol8} and it was used to demonstrate valley Hall effect for vortices \cite{sol9}.

Topological insulators with engineered potential landscapes (such as those created by arrays of waveguides or arrays of microcavity pillars) open exciting opportunity to study the impact of insulator geometry on edge currents. This problem is especially important from the point of view of demonstration of robustness of topological edge states to local perturbations/disorder and to global geometrical interface deformations. Even though many works demonstrated robustness of the edge states on sharp bends of the structure, actually only few of them depart from geometry, where two half-spaces are occupied by topologically distinct materials, and consider the case, where finite-dimensional topological insulator is fully immersed in the non-topological environment. The representative feature of the latter setting is that characteristic localization length of the edge states near the surface may be comparable with the size of the insulator. Rare examples, all of which still consider quite large structures, include finite-size topological quasi-crystals with helical waveguides \cite{helix5}, {\color{red} study coupling of edge states from the opposite boundaries in the vicinity of degeneracies in the energy spectrum in finite photonic crystal ribbons \cite{finite1,finite2})}, and narrow polaritonic ribbons, extended in only one direction \cite{rabi1}, finite-dimensional topological lasers made with photonic crystals \cite{laser1}, arrays of coupled-ring resonators \cite{laser2,laser3}, and polariton microcavity pillars \cite{laser4,laser5}, to name a few. This leaves open the important question connected with dimensionality and limits for miniaturization of topological insulators: What is the minimal size of the structured insulator sufficient for clear emergence of topological properties and allowing excitation of protected edge currents immune to structure deformations? {\color{red}This question is especially relevant for optical and optoelectronic systems, where fabrication of really extended structures is technologically challenging, and all experiments illustrating emergence of edge states and their topological protection are necessarily performed in finite geometries. To the best of our knowledge the answer to this important question has not been provided yet.}

To answer it and to demonstrate how topological properties build up with increase of dimensions of topological insulators we consider polariton insulator formed by the array of microcavity pillars with appreciable spin-orbit coupling, arranged into honeycomb structure with finite and gradually increasing number of periods along each side of the array, and placed in the external magnetic field. We find eigenmodes of such arrays, identify which of them give rise to protected edge currents, and illustrate increasing number of the topologically protected edge states with increase of the size of this structure. {\color{red}Our findings clearly show that persistent edge currents are possible already in sufficiently small structures having only several periods along each side of the array and that these currents are not destroyed by such disturbances as missing pillars. This opens the way to potential realization on the basis of such structures of compact coupling elements operating with topologically protected states.} Since polariton condensates are essentially dissipative, they require pump for their excitation that can be non-resonant \cite{laser4,nonr1,nonr2,nonr3} or resonant \cite{bist1,bist2}. In the latter case, when the pump is taken as an external forcing imprinting both energy and momenta onto microcavity polaritons, we obtain rich bistability effects that were not previously explored for finite-dimensional topological structures. We found that even trivial-phase plane wave pump can excite edge currents if it has sufficiently large projection on corresponding linear eigenmode with eigenvalue belonging to the topological gap. The pump frequency corresponding to most efficient excitation shifts with increase of pump amplitude. Vortex pump can excite topological currents too, whose direction turns out to be the same even for opposite topological charges of the pump beam.

\section{Topological properties of finite arrays}
We describe the evolution of a spinor polariton condensate in the array of microcavity pillars \cite{polar8} in circular polarization basis using coupled Gross-Pitaevskii equations for the spin-positive and spin-negative components of the polariton wavefunction $\Psi =(\psi_+,\psi_-)^\textrm{T}$ ~\cite{polar1,sol4,bist2}:
\begin{equation}
\begin{aligned}
i\frac{\partial \psi_{\pm } }{\partial T}=&-\frac{{{\hbar ^2}}}{{2m}}\left( {\frac{{{\partial ^2}}}{{\partial X^2}}+\frac{{{\partial ^2}}}{{\partial Y^2}}} \right) \psi_{\pm }+\frac{{\beta {\hbar ^2}}}{m} \left( {\frac{{{\partial}}}{{\partial X}}\mp i\frac{{{\partial }}}{{\partial Y}}} \right)^2\psi_{\mp  } \\
&+{\varepsilon}_0 [ \mathcal{R}(X,Y)\psi_{\pm } \pm \Omega\psi_{\pm }+(|\psi_{\pm }|^2 +\sigma |\psi_{\mp }|^2)\psi_{\pm } \\
&-i\gamma\psi_{\pm }+H_{\pm}(T)].
\end{aligned}
\label{eq:refname1}
\end{equation}

\begin{figure}
\centering
\includegraphics[width=\linewidth]{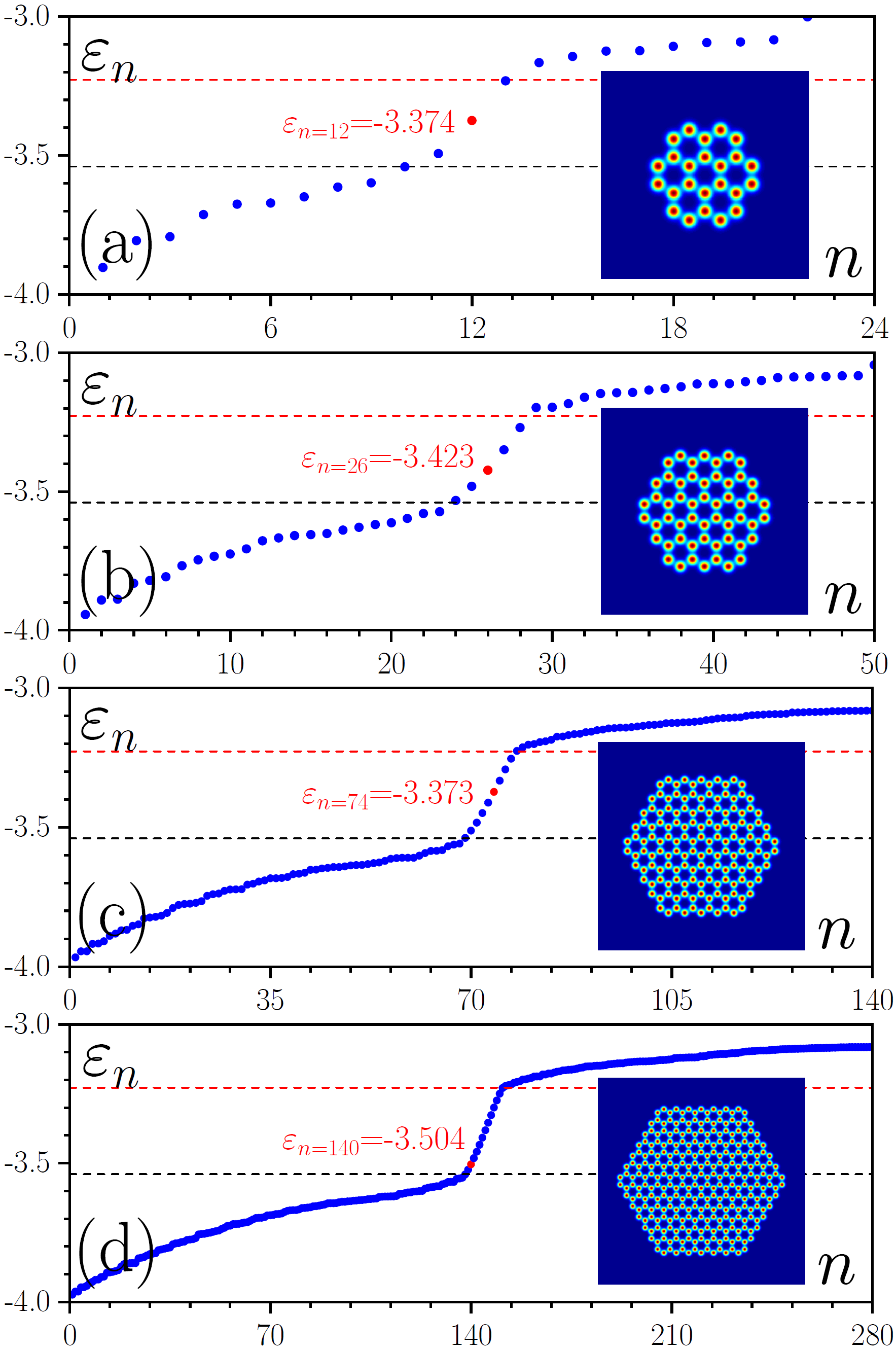}
\caption{Energies $\varepsilon_n$ of linear modes versus mode index $n$ in \color{red}{$\mathcal{H}_2$ (a), $\mathcal{H}_3$ (b), $\mathcal{H}_5$ (c), and $\mathcal{H}_7$ (d) arrays (shown in the insets). Dashed lines show borders of the topological gap in the bulk array. Red dots indicate edge modes that are excited most efficiently when pump is present. Energies of these modes are indicated too.} Here and in all figures below $\beta=0.3$ and $\Omega=0.5$.}
\label{fig1}
\end{figure}

Here, $m$ is the effective polariton mass; $\beta$ is the dimensionless parameter characterizing strength of the effective spin-orbit coupling arising from polarization-dependent tunneling between neighboring pillars in the array \cite{soc1,soc2,soc3}; $X, Y$ and $T$ are the physical distances and time, which are replaced by their normalized counterparts $(x, y, t)=(X/L, Y/L, T\varepsilon_0/\hbar)$ in our modeling; $L$ is the characteristic transverse scale; $\Omega$ is the Zeeman splitting in the external magnetic field; $\gamma$ are linear losses, which are taken identical for both spinor components; $H_{\pm}=h_{\pm}e^{-i\varepsilon T \varepsilon_0/\hbar}$ is the polarized resonant pump with dimensionless pump frequency detuning $\varepsilon$. All energy parameters in Eq. (1) are expressed through the characteristic energy $\varepsilon_0=\hbar^2/mL^2$. The potential landscape created by the array of microcavity pillars is described by the function $\mathcal{R}=\sum_{n,m} \mathcal{Q}(x-x_n,y-y_m)$, where $\mathcal{Q}(x,y)=-pe^{-[(x-x_n)^2+(y-y_m)^2]/d^2}$ stands for contribution from individual pillars of width $d$ and depth $p$. The pillars are arranged into finite hexagonal structure (see insets in Fig.\ref{fig1} ) with different number $k$ of periods along each side. One period is given by $\textrm{Y}=3^{1/2}a$, where $a$ is the separation between neighboring pillars. For convenience, we denote structures with different number of periods $k$ as $\mathcal{H}_k$, i.e. the minimal structure with one honeycomb cell is $\mathcal{H}_1$ in these notations. Assuming characteristic scale $L$ of $2~\mu \textrm{m}$ and effective mass $m=10^{-34}~\textrm{kg}$ one obtains characteristic energy scale of $\varepsilon_0=0.17~\textrm{meV}$. Further we consider the structure creating the potential landscape with $d=0.5$ ($1~\mu \textrm{m}$ radius of pillars), $p=8$ $(\sim 1.36~\textrm{meV}$ depth of the potential landscape), $a=1.4$ ($2.8~\mu \textrm{m}$ separation between pillars); and Zeeman splitting of energies of spin-positive and spin-negative polaritons of $\Omega=0.5$ $(\sim 0.085~\textrm{meV})$. We also took into account repulsion between polaritons of the same spin and weak attraction $\sigma =-0.05$ between polaritons with opposite spins. Linear losses are taken at the level of $\gamma=0.005$ - this parameter only defines the width of bistability curves and can be varied in a broad range.

\begin{figure}
\centering
\includegraphics[width=0.9\linewidth]{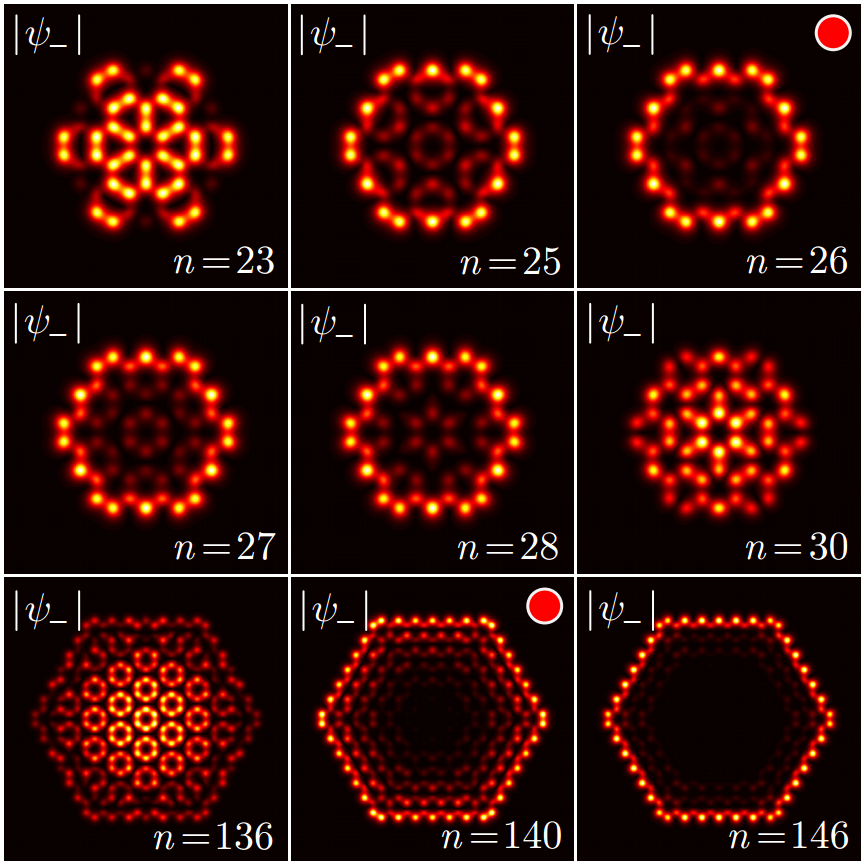}
\caption{Examples of linear modes supported by the $\mathcal{H}_3$ (first and second lines) and $\mathcal{H}_7$ (third line) structures. Modes marked with red circles are excited most efficiently in the presence of pump.}
\label{fig2}
\end{figure}

We first consider the energy spectrum of our finite-dimensional structures in the linear, pump-free and loss-free configuration {\color{red} by substituting the expression $\psi_{\pm}(x,y)=u_{\pm}(x,y)e^{-i\varepsilon t}$ for mode shapes into linear version of Eq. (1) with $\gamma=0$ and $H_{\pm}=0$ and solving the resulting linear eigenvalue problem $\varepsilon u_{\pm }=-\frac{1}{2}(\partial^{2}_x+\partial^{2}_y)u_{\pm}+\beta (\partial_x \mp i\partial_y)^2u_{\mp}+Ru_{\pm} \pm \Omega u_{\pm}$ using plane-wave expansion method}. We calculated energy spectra for a wide range of $\mathcal{H}_k$ structures, starting from $k=2$ (since for $k=1$ there is no principal distinction between bulk and edge modes). Representative examples of spectra, where all modes are sorted by increasing energy such that mode $n=1$ has lowest $\varepsilon$, for gradually increasing $k$ values are shown in Fig.~\ref{fig1}. In contrast to the semi-infinite topological insulators, where the energy spectrum is continuous and the number of states (edge states) emerging in the topological gap (the latter opens under combined action of Zeeman splitting and spin-orbit coupling breaking time-reversal symmetry of governing system of equations \cite{sol4}) is uncountable, the finite structures possess discrete energy spectra with only finite number of modes having energies within topological gap of the continuous system. For clarity, the borders of this gap are indicated by the dashed lines in Fig.~\ref{fig1}. All such modes, appearing already for $\mathcal{H}_2$ structure, have nontrivial phase distribution and they are recognized as edge states, since they have density maxima at the edges of the structure and fade away into its depth (see, for instance, the examples of such in-gap modes 25-28 from the $\mathcal{H}_3$ structure and in-gap modes 140,146 from the $\mathcal{H}_7$ structure in Fig.~\ref{fig2}). In contrast, all modes with energies outside the bulk gap (e.g., modes 23 and 30 from Fig.~\ref{fig2}) are clearly the bulk modes. The number of the edge states increases with the increase of the size of the structure, the energy spacing between different modes gradually decreases - in some sense we observe gradual transition to infinite system with increase of $k$, but the important conclusion is that even in minimal $\mathcal{H}_2$,$\mathcal{H}_3$ structures there is clear border between topological and bulk states (see Fig.~\ref{fig5} below confirming that edge states remain topologically protected even in such small structures). Interestingly, that by analogy with properties of edge states in large systems, the localization of modes in finite-dimensional insulator depends on how close the energy of the mode to the edge of the gap - compare last two panels in Fig.~\ref{fig2}, where mode 140 has energy at the border of the gap and deeply penetrates into array, while mode 146 has energy close to the gap center and is strongly localized.

\section{Bistability for plane-wave pump}
Having linear properties of the system established, we now turn to practically important question of excitation of topological currents in this finite-dimensional system. These currents can be excited using resonant pump. At unclosed straight interfaces excitation of edge currents would require matching of pump momentum with Bloch momentum of the edge state parallel to the interface. It turns out that in finite-dimensional system the excitation process is less demanding because the very configuration of the structure defines which mode will be preferentially excited by plane-wave pump with trivial phase, at the same time leaving full freedom for mode selection by pump with nontrivial, e.g. vortical phase. In the case of plane-wave pump $H_{\pm}(t)=h_{\pm}e^{-i\varepsilon t}$ nonlinear modes of the system (1) have the form   $\psi_{\pm}(x,y)=u_{\pm}(x,y)e^{-i\varepsilon t}$, where now energy $\varepsilon$ is dictated by the pump and functions $u_{\pm}$ obey

\begin{equation}
\begin{aligned}
\frac{1}{2}(\partial^{2}_x+\partial^{2}_y)u_{\pm}-\beta (\partial_x \mp i\partial_y)^2u_{\mp}-Ru_{\pm} \mp \Omega u_{\pm}\\
-(|u_{\pm }|^2 +\sigma |u_{\mp }|^2)u_{\pm }+i \gamma u_{\pm }-h_{\pm}+\varepsilon u_{\pm }=0.
\end{aligned}
\label{eq:refname1}
\end{equation}

\begin{figure}
\centering
\includegraphics[width=\linewidth]{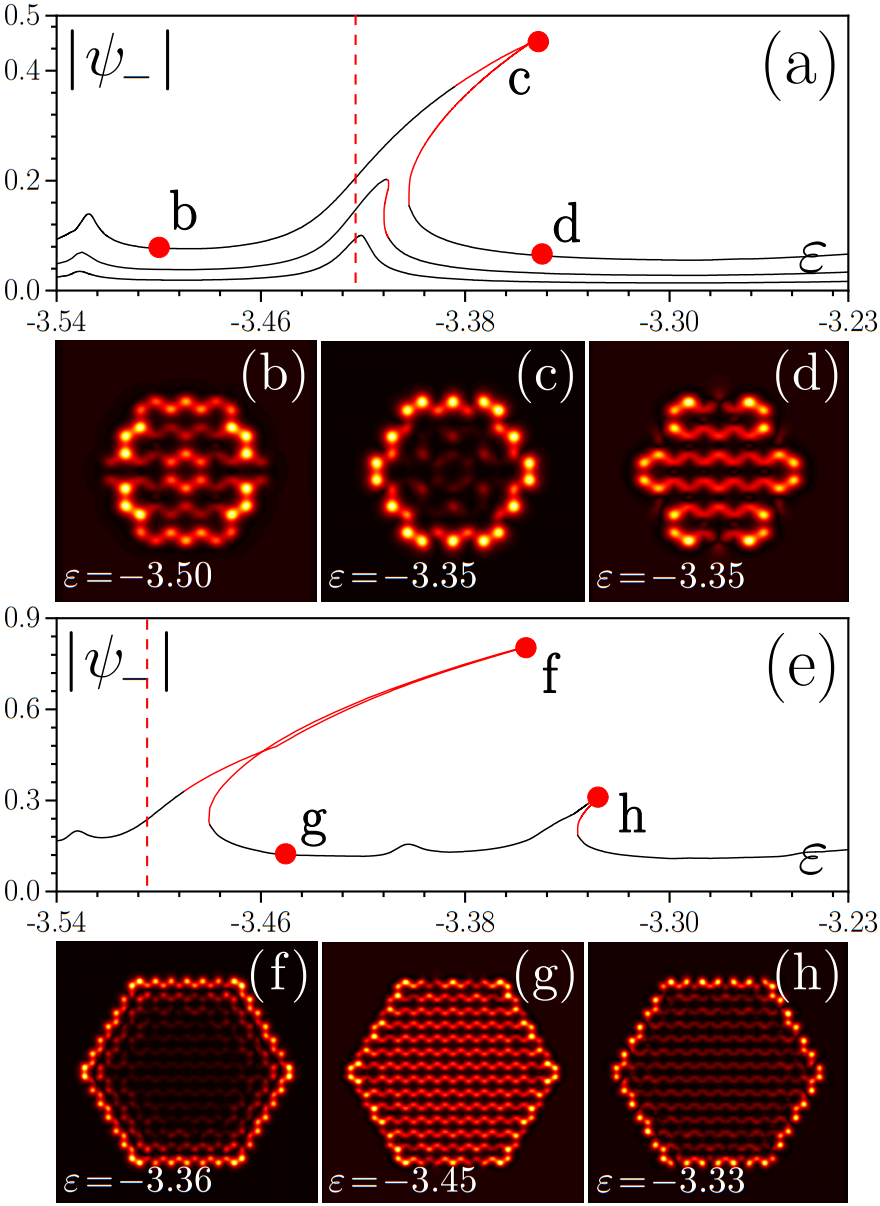}
\caption{Peak amplitude of the dominating $\psi_-$ component versus detuning $\varepsilon$ for $\mathcal{H}_3$ (a) and $\mathcal{H}_7$ (e) arrays. {\color{red}In (a) pump amplitudes $h_{\pm}=0.002$ (bottom curve), $0.004$ (middle curve), and $0.008$ (top curve) are used, in (e) $h_{\pm}=0.016$. $|\psi_-|$ distributions in (b)-(d) and (f)-(h) correspond to red dots. Namely, the modes shown are obtained at $\varepsilon=-3.50$ (point b), $\varepsilon=-3.35$ (points c and d), $\varepsilon=-3.36$ (point f), $\varepsilon=-3.45$ (point g), and $\varepsilon=-3.33$ (point h).} Stable branches are shown black, unstable ones are shown red. Dashed red lines indicate energy of linear modes that are most efficiently excited.}
\label{fig3}
\end{figure}

Eq.~(2) was solved using Newton method by fixing the pump strength (we consider a linearly polarized pump, namely, $h_+=h_-$) and scanning pump energy across the range corresponding to topological gap. Excitation of edge currents is a highly selective resonant process that is most effective when pump energy coincides with that of one of the linear modes of the structure, and it has narrow bandwidth (defined by losses $\gamma$). Typical dependencies of the amplitude of the $\psi_-$ component of nonlinear mode (that substantially exceeds amplitude of $\psi_+$ component for selected sign of Zeeman splitting) on pump energy $\varepsilon$ are shown in Fig.~\ref{fig3} for $\mathcal{H}_3$ and $\mathcal{H}_7$ structures. Despite the fact that there are multiple linear modes in the gap (see Fig.~\ref{fig1}), linearly polarized pump effectively excites only one of them in the $\mathcal{H}_3$ structure [Fig.~\ref{fig3}(a)] and two of them in the $\mathcal{H}_7$ structure [Fig.~\ref{fig3}(e)]. The energies of the most efficiently excited modes (in the linear limit) are indicated by dashed lines in Fig.~\ref{fig3}(a),(e) and by red circles in Fig.~\ref{fig1}. Such selectivity in mode excitation can be explained by calculating projections of plane-wave pump on the profiles of linear eigenmodes of the array. It turns out that due to symmetry of these modes, the projection is practically zero for most of them, except for several modes that are eventually excited. The maximal amplitude of the wave in resonance is dictated by the modulus of this projection, i.e. excitation efficiency strongly differ. The larger is the structure, the more modes are excited, but corresponding resonances are usually well-separated. For sufficiently strong pumps the resonance curves show pronounced nonlinearity-induced tilt, {\color{red}sometimes leading to bistability as in Fig.~\ref{fig3}(a)}, and may even {\color{red}form loops}, like the one seen in the left resonance in Fig.~\ref{fig3}(e). Notice that far from the resonance the nonlinear state represents a mixture of several modes and has considerable amplitude in the bulk of the array, but close to the resonance tip the nonlinear state clearly localizes at the edge, see Fig.~\ref{fig3}(b)-(d) and Fig.~\ref{fig3}(f)-(h) corresponding to red dots on corresponding bistability curves.

\begin{figure}
\centering
\includegraphics[width=\linewidth]{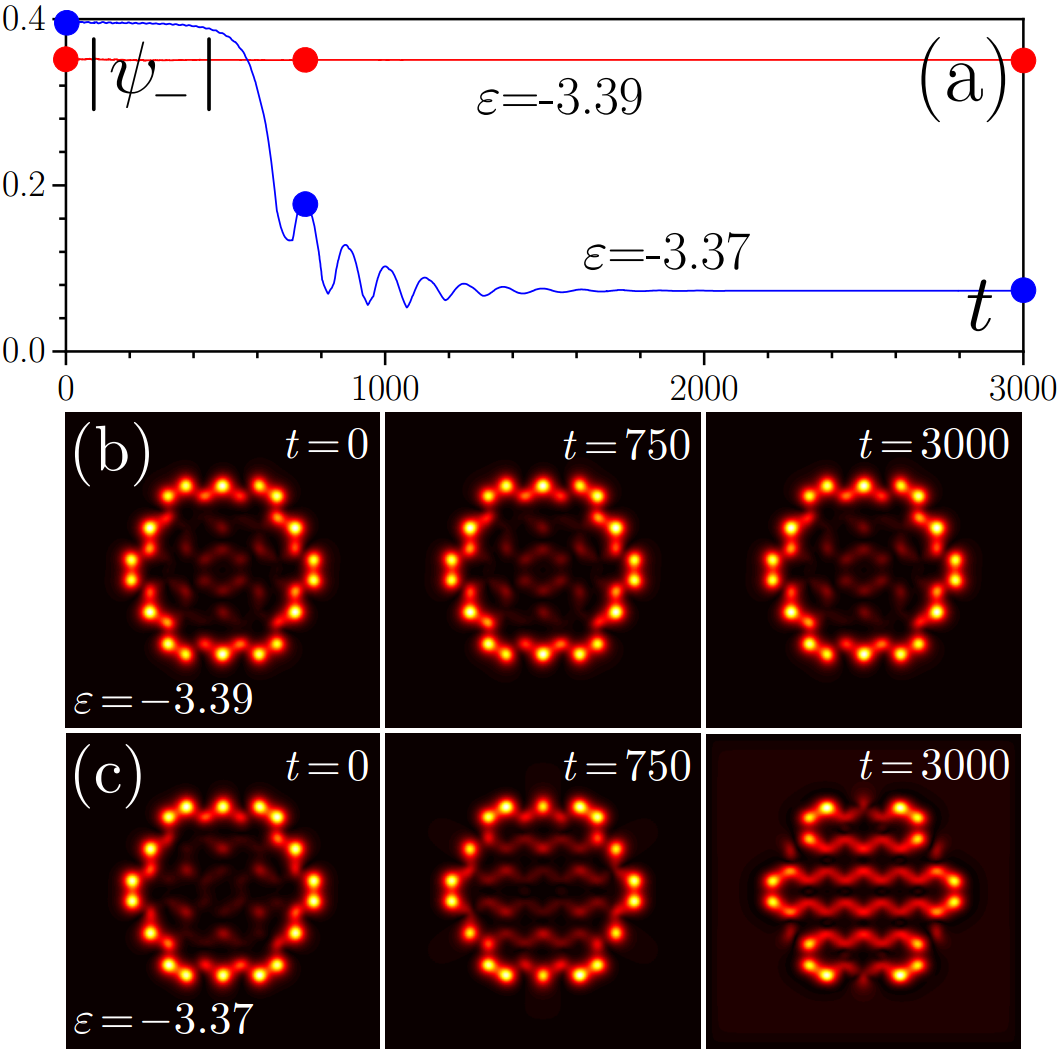}
\caption{Peak amplitude of $\psi_-$ component versus time (a) and examples of spatial $|\psi_-|$ distributions in different moments of time [corresponding to dots in (a)] showing stable evolution at $\varepsilon=-3.39$ (b) and unstable evolution at $\varepsilon=-3.37$ (c) of the perturbed edge states from the upper branch of the
resonance in $\mathcal{H}_3$ array at $h_{\pm}=0.008$.}
\label{fig4}
\end{figure}

{\color{red} We studied stability of these nonlinear modes by direct numerical integration of Eq.~(1) with the initial conditions $\psi_{\pm}=u_{\pm}(1+\alpha_{\pm})$, where  $u_{\pm}$ is the function describing edge state and $\alpha_{\pm}$ is the broadband small-amplitude noise. Corresponding modes were allowed to evolve up to times $t\sim 10^4$ greatly exceeding polariton lifetime. The nonlinear mode is considered stable if the mode maintains its waveform over time interval $t\sim 10^4$, and unstable otherwise.} Stable and unstable branches detected with this method are indicated in Fig.~\ref{fig3}(a),(e) with black and red colors, respectively. For sufficiently large pump amplitudes only part of the upper branch is stable - corresponding waves belonging to stable region are well-localized at the edge. One example of stable evolution and corresponding profiles of initially perturbed nonlinear mode are shown in Fig.~\ref{fig4}(a),(b). The modes from the unstable part of the upper branch usually jump to the lower branch on bistability curve, this is accompanied by extension of the mode into bulk of the array, see Fig.~\ref{fig4}(a),(c). Lower branches are always stable, and middle branches are unstable. Interestingly, in the $\mathcal{H}_7$ structure middle and upper branches may exchange their order, since they form loop [Fig.~\ref{fig3}(e)], with both branches being unstable.

\begin{figure}
\centering
\includegraphics[width=0.9\linewidth]{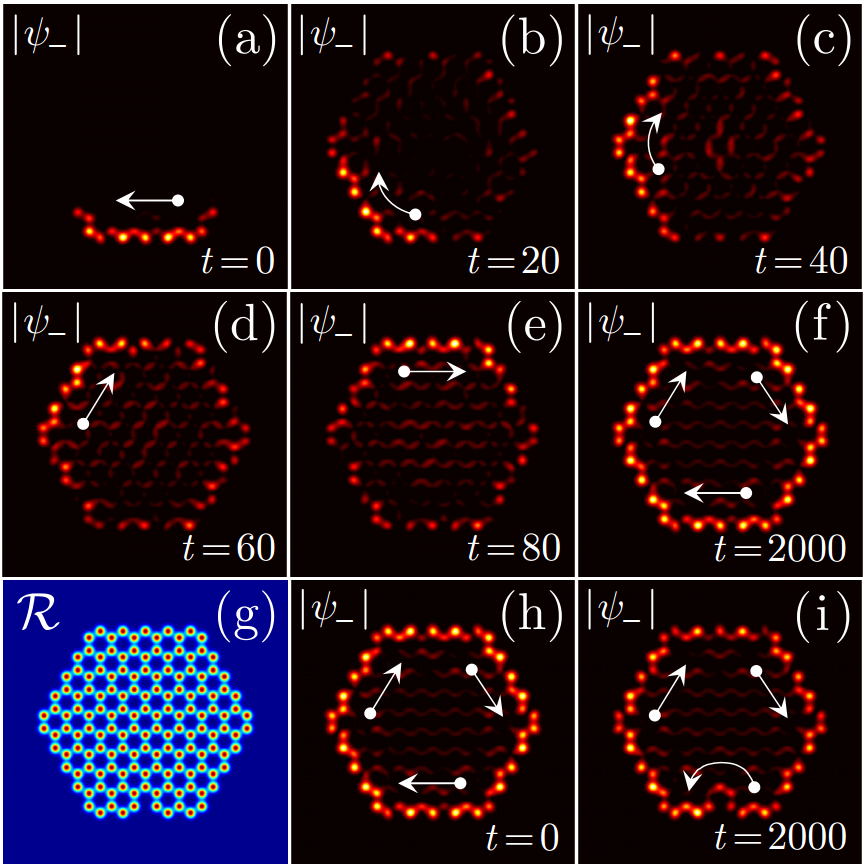}
\caption{(a)-(f) Example of circulation of localized input in the $\mathcal{H}_5$ array that ends up with excitation of stable edge state. (h),(i) Evolution of edge state in the $\mathcal{H}_5$ array with defect, shown in (g). In all cases $h_{\pm}=0.016$, $\varepsilon=-3.345$.}
\label{fig5}
\end{figure}

We emphasize that despite the fact that arrays considered here are very small in comparison with structures typically used to illustrate topological excitations, the modes supported by such arrays are topologically protected entities. To demonstrate unidirectional nature and robustness of such excitations we imposed localized envelope on one of them and let it evolve. Fig.~\ref{fig5}(a-f) show how truncated nonlinear edge mode gradually evolves into edge state that extends over entire boundary. Circulation in the clockwise direction indicated by white arrows is also obvious at the initial stages of evolution. Notice that circulation direction is defined exclusively by the sign of spin-orbit coupling coefficient $\beta$ and sign of Zeeman splitting $\Omega$ defined by the direction of the external magnetic field applied to the structure, i.e. the same circulation direction was observed for all excited modes from topological gap and for any size of the $\mathcal{H}_k$ structure, confirming unidirectional character of edge states in this system that is not affected by the pump. Reversal of the sign of $\Omega$ leads to reversal of edge currents. {\color{red}Edge states in finite-dimensional arrays are also topologically protected and insensitive to the defects in the underlying structure. One example is shown in Fig.~\ref{fig5}(g)-(i), where initially unperturbed edge mode from the stable branch of the resonance curve gradually reshapes to accommodate to surface defect in the form of missing pillar.} The reshaping is local and it only affects the wave profile around the defect, since peak amplitude of the state remains unchanged at large times.

\section{Bistability for vortex pump}
\begin{figure}
\centering
\includegraphics[width=\linewidth]{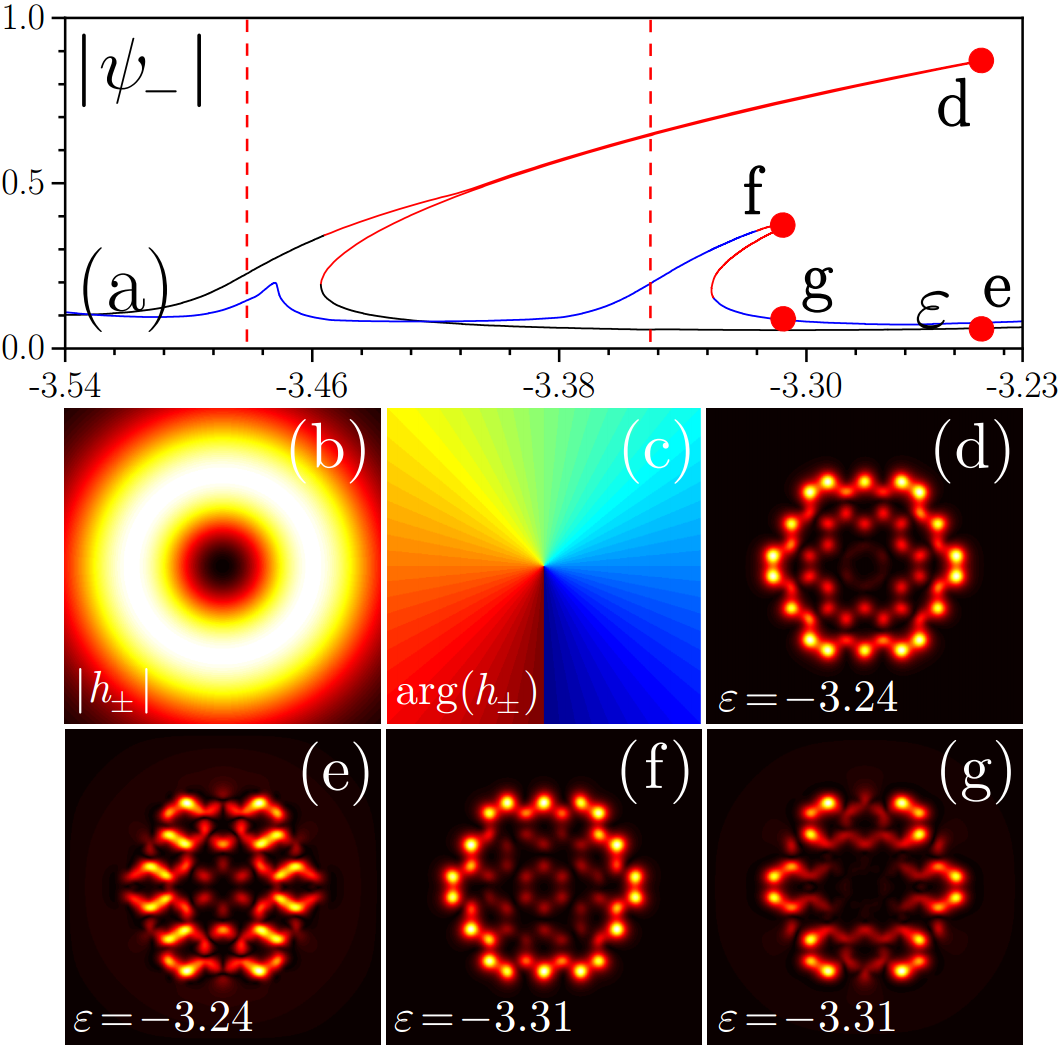}
\caption{(a) Peak amplitude of the dominating $\psi_-$ component versus detuning $\varepsilon$ for vortical pump with $m=+1$ (black-red curves) and $m=-1$ (blue-red curves) in $\mathcal{H}_3$ structure. Pump amplitude is 0.02 and width is 8. Stable branches are shown black and blue, unstable ones are shown red. Pump modulus (b) and phase (c) for $m=+1$. $|\psi_-|$ distributions for $m=+1$ (d),(e) and $m=-1$ (f),(g) pump correspond to red dots in (a). Dashed red lines show energies of linear modes most efficiently excited by the vortex pump: $\varepsilon=-3.481$ ($m=+1$ pump) and $\varepsilon=-3.350$ ($m=-1$ pump).}
\label{fig6}
\end{figure}

As it was shown above, plane-wave pump excites only certain specific modes of the structure that exhibit clockwise circulation due to broken time-reversal symmetry. The pump with nontrivial phase distribution, such as vortical one, brings additional rich possibilities for control of the excitations in finite-dimensional insulators. Here we are mainly interested how topological charge $m$ of the pump $h_{\pm}=h_0(r/w)^{|m|}e^{-r^2/w^2+im \phi}$ affects emerging edge states. It is well-known that in nontopological polariton systems resonant pump imposes its phase on the condensate, so the question arises whether similar phenomenon exists in topological systems. It turns out that the response of topological insulator to vortex pump is completely different. Fig.~\ref{fig6}(a) compares resonance curves for pumps with opposite topological charges. They are markedly different: the resonance for $m=+1$ pump is much stronger than that for $m=-1$ pump and they in fact occur at two different energy values, i.e. different topological modes are excited. The profile of such vortical pump is shown in Fig.~\ref{fig6}(b),(c), while profiles of excited modes in tips of two resonances are shown in Fig.~\ref{fig6}(d) and (f). Both these modes are clearly localized at the edge of the structure. Because topological system admits only clockwise currents for selected sign of $\beta,\Omega$, both modes reveal circulation in the clockwise direction, when they are truncated. It should be stressed that the structure of resonances is different for other values of topological charge $m$ of the pump, i.e. by changing $m$ one can excite any desired nonlinear edge state available in this structure. The velocity of circulation in corresponding edge states will be determined not only by the modulus of topological charge of the pump, but also by its sign. We also studied stability of edge states excited by vortex pump. As before unstable branches are shown with red color in Fig.~\ref{fig6}(a), while stable branches are shown black/blue. If pump is not too large the entire upper branch may be dynamically stable (see right resonance).

\section{Summary}
Summarizing, we predicted that even small-size polariton topological insulators immersed into non-topological environment demonstrate clear unidirectional edge currents that are not destroyed by structure defects. The practical system that we addressed admits very selective excitation of edge currents by pump with nontrivial phase distribution. Robustness of topological effects together with micron-scale dimensions of structures considered here, make them promising for construction of practical topological devices. They can be fabricated with different shapes and sizes offering flexibility not available in extended periodic structures and opening new prospects for realization of topological switching and transmission devices, resistant to perturbations. The results reported here may be directly extended to other optical or matter-wave systems, such as spin-orbit coupled BEC or arrays of helical waveguides, where topological protection is guaranteed by broken time-reversal symmetry.
\begin{acknowledgement}
Y.V.K. acknowledges funding of this study by RFBR and DFG according to the research project No.18-502-12080. F. Y. acknowledges support by Natural Science Foundation of Shanghai (No.19ZR1424400) and NSFC (No.61475101). 
\end{acknowledgement}

%

\end{document}